\documentclass[12pt]{article}
\usepackage[a4paper, total={7in, 10in}]{geometry}
\usepackage[parfill]{parskip}
\usepackage{jhep-mod}
\usepackage{bm}
\usepackage{soul}
\usepackage{amssymb,amsmath,amsthm}
\usepackage{mathrsfs}
\usepackage[utf8]{inputenc}
\usepackage{enumerate}
\usepackage{bigints}
\usepackage{xcolor}
\usepackage{appendix}
\usepackage{graphicx}
\usepackage{float}
\usepackage{tikz}
\usepackage{setspace}
\usepackage{cancel}
\usepackage{doi}
\definecolor{purple}{rgb}{1,0,1}
\definecolor{lime}{HTML}{A6CE39} 

\definecolor{lime}{HTML}{A6CE39}
\newcommand{\orcidicon}{%
	\begin{tikzpicture}
	\draw[lime, fill=lime] (0,0) 
		circle [radius=0.16] 
		node[white] {{\fontfamily{qag}\selectfont \tiny ID}};
	\draw[white, fill=white] (-0.0625,0.095) 
		circle [radius=0.007];
	\end{tikzpicture}
	\hspace{-5mm}
}
\newcommand\orcidMatt{{\href{https://orcid.org/0000-0003-1088-6485}{\orcidicon}}}

\begin{document}
\title{\Huge Which number system is ``best''\\
 for describing empirical reality?}
\author{Matt Visser\!\orcidMatt\!}
\affiliation{School of Mathematics and Statistics\\
Victoria University of Wellington, PO Box 600, \\
Wellington 6140, New Zealand}
\emailAdd{matt.visser@sms.vuw.ac.nz}
\abstract{
\vspace{1em}

Eugene Wigner's much-discussed notion of the ``unreasonable effectiveness of mathematics'' as applied to describing the physics of empirical reality is simultaneously both trivial and profound. After all, the relevant mathematics was (in the first instance) originally developed in order to be useful in describing empirical reality. On the other hand, certain aspects of the mathematical superstructure have by now taken on a life of their own, with at least \emph{{some}} features of the mathematical superstructure greatly exceeding anything that can be directly probed or verified, or even justified, by empirical experiment. Specifically, I wish to raise the possibility that the real number system (with its nevertheless pragmatically very useful tools of real analysis and mathematically rigorous notions of differentiation and integration) may nevertheless constitute a ``wrong turn'' (a ``sub-optimal'' choice) when it comes to modelling empirical reality. Without making any definitive recommendation,  I shall discuss several reasonably well-developed alternatives.



\vskip 10 pt
\noindent 
V1: 26 December 2012; (arXiv 1212.6274 math-ph); V2: 16 September 2022; \\\LaTeX-ed \today.

\vskip 10 pt
\noindent 
Extended version of an essay originally written for the FQXi 2012 essay contest: \\``Questioning the foundations. Which of our basic physical assumptions are wrong?''
Now further extended with significant additional commentary and references.

\bigskip
\noindent
This version published as MDPI Mathematics  {\bf10 \# 18} (2022) 3340; \\\doi{10.3390/math10183340}

}


\vspace{2pc}
\noindent

\maketitle

\markboth{Which number system is ``best'' for describing empirical reality?}{ }
\clearpage
\markboth{Which number system is ``best'' for describing empirical reality?}{}
\def\d{{\mathrm{d}}}
\def\O{{\mathcal{O}}}
\def\omicron{o}
\def\CC{\mathbb{C}_0}
\def\N{\mathbb{N}}
\def\Z{\mathbb{Z}}
\def\Q{\mathbb{Q}}
\def\R{\mathbb{R}}
\def\C{\mathbb{C}}
\section{Introduction}\label{S:intro}

Over 60 years ago, Eugene Wigner introduced and developed the idea of the ``unreasonable effectiveness of mathematics''~\cite{Wigner}  when it comes to  describing the physics of empirical reality---this is an extremely tricky issue that bears repeated re-examination.  There are definitely very many powerful assumptions built into modern mathematics, and specifically that part of mathematics used for investigating physics, and only some of these assumptions can be asserted to be solidly based on direct empirical input. 

While the many and varied successes of modern physics clearly indicate that we are doing \emph{something}  right, the very fact that there is also a long list of things that we do \emph{not} know how to do (both in physics and mathematics) suggests that we might wish to double-check for possible ``wrong turns'' (sub-optimal choices) in the development of the \emph{status quo}.  

In particular, some commentators have been known to bemoan the relative ``unreasonable \emph{ineffectiveness} of mathematics'' when it comes to dealing with the biological and social sciences. This may more be attributable to a mismatch of specific mathematical tools to specific scientific problems, rather than any deep issue with mathematics itself. 
Qualitative tools \emph{versus} quantitative tools often require significantly different mathematical frameworks. Herein, I shall explicitly elaborate on the central role of mathematical number systems, emphasizing the difference between the ``discretium'' and the ``continuum'', 
and explore the choice of  which number system is ``best'' for describing empirical reality.

\section{Pre-Calculus}\label{S:precalculus}

When studying pre-calculus mathematics, one encounters an extremely natural progression of number systems:
\begin{itemize}
\item Counting numbers, $\CC=\{1,2,3,\dots\}$.
\item Natural numbers, $\N = \{0,1,2,3,\dots\}$.
\item Integers, $\Z=\{\dots,-3,-2,-1,0,+1,+2,+3,\dots\}$.
\item Rational numbers, $\Q = \{p/q; p\in \Z, q\in \CC\}$.
\item Real numbers, $\R$.
\item Complex numbers, $\C$.
\end{itemize}

The first three stages in this progression (counting numbers, natural numbers, and integers) are extremely closely tied to empirical reality, essentially just being mathematical formalizations of the physical notion of the existence of discrete and separate objects in the observable universe.  (Though even then, the implicit notion of (countable) infinity hiding in these number systems might give some people pause for concern.)

However, already once one moves to the rational numbers (while no-one should dispute the usefulness of the resulting mathematics), there is a very real and nontrivial question as to how closely they model empirical reality. Since the rational numbers are ``infinitely divisible'', if one uses the rationals to model space and time, then one is already making very strong physical assumptions regarding the empirical physical structure of space-time, effectively, that there is an infinitely divisible space-time. 

\enlargethispage{20pt}
Viewed in hindsight (and with a modern change of terminology) that no one was too worried about this issue for several millennia can be inferred from the fact that Euclid's axioms~\cite{Euclid} of geometry explicitly allowed for ``infinitely divisible'' line segments (and so implicitly allowed, at the very least, what might now be called ``rational-coordinate Cartesian geometry'').
Of course, Euclid's axioms assert much more than this, quite naturally leading to number systems considerably more complicated than the rationals, but still simpler than the reals---and I will have much more to say about this below.

\vspace{-5pt}
\section{Calculus}\label{S:calculus}
\vspace{-5pt}

Much more recently, somewhat over 300 years ago, Newton~\cite{Newton1,Newton2} and Leibniz~\cite{Leibniz1,Leibniz2} developed the differential and integral calculus as general purpose mathematical tools~\cite{note}. (The calculus was developed largely for calculating slopes and volumes---thereby extending the ``method of exhaustion'' developed by the ancient Greeks~\cite{exhaustion1, exhaustion2}; and also for the extremely practical purpose of predicting planetary orbits). Putting these pragmatic tools on a rigourous footing (as Berkeley asked: ``What, pray tell, are these \emph{differentials}, these ghosts of departed quantities?''~\cite{Berkeley}), took well over a century, and led Dedekind, Cauchy, Weirstrauss, and numerous others to develop and codify the real number system---with its associated theory of real analysis and mathematically rigourous theories of differentiation and integration~\cite{Hardy, dirichlet-series, Shilov}. 

No one can reasonably doubt the pragmatic effectiveness (the possibly ``unreasonable effectiveness'') of real analysis~\cite{Wigner}---just consider the theory of ordinary and partial differential equations, the associated existence and uniqueness results, and the vast body of mathematical physics that has grown up over the last few centuries---almost all of which is based on the straightforward application of real analysis (or its offspring, complex analysis). \enlargethispage{40pt}

However, and this is a very big however, the very successfulness of real analysis relies upon the fact that the real number system is a mathematical idealization (not a physical model) that is by that stage very far removed from empirical reality of how one might actually make measurements, and so is very far removed from the  experimental verifiability. 
How, pray tell, might one go about experimentally verifying the implicit claim underlying the use of real analysis in physics---the implicit claim that the measurements made by physical clocks and rulers are adequately characterized by the real number system?  (And more subtly, the often implicitly made but highly debatable assumption that one needs the \emph{entire} real number system to adequately characterize physical measurements.)

\clearpage
\section{Betwixt the Rationals and the Reals}\label{S:betwixt}

Closely related to this question is the fact (typically ignored in the training of most physicists and even many mathematicians) that there is a whole hierarchy of ``intermediate'' number systems lying between the rational numbers and the real numbers, and that it might be that one of these other number systems has more claim to be operationally relevant to physics.

Some of the other ``intermediate'' number systems are:
\begin{itemize}
\item The field of quadratic surds.

(Numbers of the form $q_1 + \sqrt{q_2}$, where $q_1$ and $q_2$ are rational).

\item Euclidean constructible numbers.

(Numbers constructible in Euclidean geometry using straight-edge and compass).

This corresponds to the set of numbersthat can be generated from the integers by a finite number of additions, subtractions, multiplications, divisions, and square root~extractions.

\item Post-Euclidean constructible numbers---I.

(Numbers constructible in Euclidean geometry using \emph{marked-edge} and compass).

\item Post-Euclidean constructible numbers---II.

(Numbers constructible in Euclidean geometry using straight-edge, compass, and \emph{rolling marked circles}).

\item Post-Euclidean constructible numbers---III.

(Numbers constructible in Euclidean geometry using \emph{marked-edge}, compass, and \emph{rolling marked circles}).

\item Algebraic numbers.

(Numbers that are the roots of integer coefficient polynomials).

\item Computable numbers (recursive numbers, computable reals).

(Computable numbers $c$ can be approximated to any arbitrary specified rational accuracy $\epsilon$ by a rational number $q$ generated by some computable function;  that is, $q=f(c,\epsilon)$ with $|c-f(c,\epsilon)|<\epsilon$).

\item Definable numbers (first-order definable without parameters).

(Definable numbers are set theoretically definable as the quantities that make some set-theoretic function ``true'').

\end{itemize}

These particular ``intermediate'' number fields all share the property of being countable (denumerable); while they all extend the rationals, their cardinality is the same as the rationals, and by Cantor's diagonalization argument, their cardinality is less than that of the reals---in this sense that they are closer to the rationals than the reals. In this sense, the possible use of such number systems would be an aspect of the great modern divide between ``discrete'' and ``continuum'' mathematics; in physics language, consider the ``discretium'' versus the ``continuum''. 

\clearpage
One might hope (though at this stage it is little more than a rather pious hope) that the reduced cardinality might in some way simplify life. For example, lattice quantum field theory (QFT) (even in the infinite volume limit), still has denumerable degrees of freedom, and is much better behaved than continuum QFT~\cite{Lattice1,Lattice2,Lattice3}. Lattice QFT can be given a much more rigourous and even non-perturbative interpretation. In contrast, in the (3+1) continuum, there are as yet no \emph{rigorously established} non-trivial (interacting) Lorentz-invariant QFTs, though perturbation theory seems to work very nicely~\cite{QFT}.

\section{The ``Best'' Number System for Physics?}\label{S:best}

Could any of these number systems be ``better'' than the reals when it comes to doing physics?  Are any one of them clearly superior to the others? Ultimately, this is an entirely pragmatic question that depends on various trade-offs---in particular, how much of standard real analysis survives in these smaller number systems?  Does one have, or can one construct, good notions of integration and differentiation? Alternatively, instead of the usual theory of differential equations, can one at the very least develop a coherent theory of finite difference equations over these number systems? 

Overall, the situation is very much less than clear, and I will not provide any specific definitive answers---there are several abstract mathematical attempts at developing alternatives to real analysis, but I will leave technical details as an exercise for one's favourite search engine---at this stage, I wish merely to raise suitable  questions, and to encourage some thought along these lines. Specifically, there is a certain lack of clarity as to which of these number systems to focus on, and the physics community might most profitably contribute to the discussion by developing some consensus as to the minimum requirements for a physically acceptable number system.

\subsection{Rational Cartesian Geometry}\label{SS:rational}

For instance, it is periodically mooted that one could get away with merely using the rational numbers as coordinates---but there is an extremely high price to pay for this, which is maybe just  too high---distances between generic points in $\Q^n$ are typically not rational, but are instead square roots of rational numbers. Under rotation, if one wishes rational coordinates to remain rational, then one can only rotate through angles such that $\sin(\theta)$ and $\cos(\theta)$ are both rational. Use the classic result that the Pythagorean triples of integers (describing the sides of right-angled triangles) are completely specified {by:} 
\begin{equation*}
(m_1^2+m_2^2); \qquad (m_1^2-m_2^2); \qquad 2 m_1 m_2; \qquad m_1,m_2 \in \N.
\end{equation*}
Then the allowable rotation angles are constrained by either:
\begin{equation*}
\sin\theta = {m_1^2-m_2^2\over m_1^2+m_2^2}; \qquad \cos\theta =  {2m_1m_2\over m_1^2+m_2^2}; 
\qquad m_1,m_2 \in \N,
\end{equation*}
or alternatively,
\begin{equation*}
\sin\theta =  {2m_1m_2\over m_1^2+m_2^2}; \qquad \cos\theta = {m_1^2-m_2^2\over m_1^2+m_2^2}; 
\qquad m_1,m_2 \in \N.
\end{equation*}

Thus, in this model, allowable rotation angles are discretized, in a very specific way, with a denumerable set of allowable rotation angles. 
Similarly, if one wishes Lorentz transformations to map rational coordinates to rational coordinates, then both $\gamma$ and $\gamma\beta$ must be rational, which is equivalent to both $\beta$ and $\sqrt{1-\beta^2}$ being rational. Consequently,~either:
\begin{equation*}
\qquad
\beta = {m_1^2-m_2^2\over m_1^2+m_2^2}; \qquad \sqrt{1-\beta^2} =  {2m_1m_2\over m_1^2+m_2^2}; \qquad
\gamma\beta =  {m_1^2-m_2^2\over 2m_1m_2}; \qquad m_1,m_2 \in \N,
\end{equation*}
or alternatively, 
\begin{equation*}
\qquad
\beta =  {2m_1m_2\over m_1^2+m_2^2};  \qquad \sqrt{1-\beta^2} = {m_1^2-m_2^2\over m_1^2+m_2^2}; \qquad
\gamma\beta =  {2m_1m_2\over m_1^2-m_2^2};
\qquad m_1,m_2 \in \N.
\end{equation*}

Thus, in this model, the allowable velocities are discretized, in a very specific way, as a denumerable set of rational fractions of the speed of light.  This is certainly an interesting set of constraints, but it must be emphasized that even with these constraints, proper distances and proper time intervals between events are typically not rational numbers, though they are at worst square roots of rational numbers.

In counterpoint, if one naively asserts that all lengths (or more precisely length ratios) are describable by rational numbers, then one already has excluded the $\{1,1,\sqrt{2}\}$ right-angled triangle from the physical universe, and for $n\in \CC$, has also excluded $\{1,n,\sqrt{n^2+1}\}$ right-angled triangles  from the physical universe. This observation (suitably repackaged) essentially goes back to the ancient Pythagorean proof of the irrationality of $\sqrt{2}$, and does seem to suggest that (if one first accepts rational number distances as being paramount) there are significant constraints on the allowable coordinates, the coordinates can at best be a subset of the rationals. Specifically, in $\Q^2$, the classic result regarding Pythagorean triples implies coordinate differences would be constrained by:
\begin{equation*}
\Delta x = q_0(m_1^2-m_2^2); \qquad \Delta y = 2 q_0 m_1 m_2; \qquad q_0 \in \Q, \quad m_1,m_2 \in \N,
\end{equation*}
with the corresponding distances being:
\begin{equation*}
d(\Delta x,\Delta y) = \sqrt{\Delta x^2 +\Delta y^2} = q_0 (m_1^2+m_2^2);  \qquad q_0 \in \Q, \quad m_1,m_2 \in \N.
\end{equation*}
Overall, these considerations raise some problematic issues, which suggest that looking at a number system somewhat beyond the rationals would be a good idea. 

\subsection{Quadratic-Surd Cartesian Geometry}\label{SS:surd}
If the coordinates are restricted to be arbitrary quadratic surds,
\begin{equation*}
x =  p + \sqrt{q}; \qquad  p,q\in\Q,
\end{equation*}
then the situation is still quite similar to that for rational coordinates. Rotation angles would be restricted by the fact that both $\sin(\theta)$ and $\cos(\theta)$ would be forced to be quadratic surds. For Lorentz transformations, both $\beta$  and $\sqrt{1-\beta^2}$ would be forced to be quadratic surds. Distances (including proper distances and proper times) would be given by square roots of quadratic surds, which generally are not themselves quadratic surds. The fact that the set of quadratic surds is not closed under the extraction of square roots is the obstruction. Overall, these considerations raise some problematic issues, which suggest that looking at a number system somewhat beyond the quadratic surds would be a good idea.

\subsection{Constructible Cartesian Geometry}\label{SS:constructible}
In contrast, the set of constructible numbers is closed under the square root operation.
Consequently, in (any version of) ``constructible number Cartesian geometry'', where coordinates are constrained to be  constructible numbers, the distances between points (including proper distances and proper times) are constructible.
Rotation angles would be restricted by the fact that both $\sin(\theta)$ and $\cos(\theta)$ would be forced to be constructible. For Lorentz transformations both $\beta$  and $\sqrt{1-\beta^2}$ would be forced to be constructible. 
This would seem to focus attention on the constructible numbers as a particularly interesting number system to explore.  

\subsection{Euclidean vs. Post-Euclidean Constructible Cartesian Geometry}\label{SS:constructible}

However, even with the constructible numbers, there are choices to be made---one has to decide precisely which construction techniques are allowable. The classic Euclidean straight edge and compass is merely the most straightforward option.
\begin{itemize}
\item 
For instance, with \emph{marked edge} and compass (as opposed to \emph{straight edge} and compass), by using a \emph{neusis construction} it is well-known that one \emph{can} trisect arbitrary angles~\cite{marked-trisect}, and also double the cube~\cite{marked-cube}. (One could also use Origami techniques to trisect arbitrary angles and to double the cube~\cite{origami}.) 
\item
In a similar vein, if one is allowed to \emph{mark} the edge of a circle, and \emph{roll} it along a straight line (thus employing a surveyor's wheel/hodometer/waywiser), then constructing two line segments whose lengths are in the ratio $\pi$ is trivial. Such a procedure is implicit in Archimedes' theorem that the area of a circle equals the area of a right angled triangle whose base is the circumference of the circle and whose height is the radius of the circle. Once this preliminary step is done, then one certainly \emph{can} square the circle.  The same effect could be obtained by wrapping a piece of string half way round a circle, marking the ends, and then straightening out the string.
\item
One could do both: In addition to a compass, use both  a \emph{marked edge} and a \emph{marked rolling circle}.
\end{itemize}

The point is that exactly which set of constructible numbers one obtains very much depends on a precise specification of the allowed construction techniques. (Thankfully these various constructible number systems are all closed under the extraction of square roots). Note that while there is no a priori explicit bound on the number of operations used in constructing constructible numbers, it is asserted that the number of operations be \emph{finite}---so at least in principle, the constructible numbers are accessible to an (arbitrarily precise) experiment. 

\section{Empirical Guidance?}\label{S:empirical}

Perhaps more operationally, any real physical experiment can be viewed as a finite bounded-resource algorithm applied to the universe that returns some finite-precision result; and any finite-precision number can be viewed as a rational number. This point of view would suggest the relevance of the \emph{computable number system}, or possibly some sub-sector thereof. (One may wish to think carefully as to whether all computable functions can be simulated by physical experiments, or whether physical experiments only probe a subset of the computable functions. One might also wish to think carefully on the ultimate limits of precision of the experimental method. For instance, while the \emph{event horizons} of general relativity are mathematically precise definitions useful for proving rigorous mathematical theorems, no finite-size finite-duration physical observer can ever detect an \emph{event horizon}, as opposed to an \emph{apparent horizon}~\cite{observability}.)

Of course there are yet other possibilities: 
\begin{itemize}
\item As very briefly outlined above, one could try to construct yet more number systems  ``between'' the rationals and the reals.

\item One could instead consider mathematical systems less powerful than the rationals. 
\begin{itemize}
\item If one is deeply offended by any notion of infinity, then even a countable number system might be too much to contemplate---one could always resort to finite fields (Galois fields~\cite{Galois}), but for empirical reasons one would have to choose an awfully big finite field. In particular,
since the ratio of the Planck length to the Hubble scale is approximately $10^{60}$,  each spacetime coordinate would presumably lie in some number system with approximately $10^{60}$ (or more) distinct elements.

\item A modification of the notion of Galois fields is provided by the \emph{residue number systems} often used in cryptographic applications.
In \emph{residue number systems} one works within the set of integers and chooses a finite set of co-prime moduli $\{m_1,m_2, \cdots, m_k\}$ and represents the number of interests 
by a finite set of residuals $\{r_1,r_2, \cdots, r_k\}$ with $0\leq r_i < m_i$. 
This construction can represent at most $M= \prod_{i=1}^k m_k$ distinct numbers.

\item If one even objects to using mathematical number fields, then  there are always finite geometries to play with, or more abstractly finite matroids~\cite{matroids}.

(With, four spacetime dimensions,  approximately $(10^{60})^4 = 10^{240}$ or more distinct points or elements respectively---though the very concept of a finite matroid with more than $10^{240}$ distinct elements is likely to seriously perturb several of my colleagues.)

\item Finite cellular automata, because they have only a finite number of internal states, can only produce a finite number of outputs; so finite cellular automata  are definitely ``weaker''  than the rationals. In counterpoint infinite cellular automata are a bit of a wild card~\cite{vonNeumann, Ulam, Denby:1987, Meyer:1996, Bialynicki-Birula:1993, Kadanoff:1989, Farrelly:2019}. Some infinite cellular automata are known to be Turing complete, and can certainly produce any rational number as output. The generic situation is still somewhat uncertain, but because infinite cellular automata by construction have, at most, a denumerable number of internal states, they inherently deal with the discretium, not the continuum~\cite{zipf,coarse,infinite}.

\item Other number systems weaker than rationals include all the finite-precision number systems used in numerical computing. These include:

\begin{itemize}
\item Fixed-point arithmetic: Numbers of the form 
$\{ x = (\hbox{integer})/(\hbox{fixed integer})\}$, where the integers in question are taken to be finite precision.  This provides a finite subset of the rationals. 

\item Floating-point arithmetic:
Numbers of the form $\{ x = (\hbox{integer})\times(\hbox{base})^{(\hbox{exponent})} \}$,
where the integer and exponent are again taken to be finite precision.  This provides a finite subset of the rationals. 
\end{itemize}

Both of these number systems can prove very useful in implementing numerical algorithms, which by their very nature must be approximate. 
A variant of these finite precision number systems is the use of \emph{interval arithmetic} where one computes upper and lower bounds to the physical quantity of interest. 
\end{itemize}

\item In the other direction, one could go to extremes even larger than the reals:
\begin{itemize}
	
\item The complex numbers (and in particular holomorphic functions) have their own extremely tantalizing aspects. Sometimes complex analysis can be extremely useful, and complex manifold theory is fascinating~\cite{Complex:1, Complex:2,Flaherty}. However, it is less than clear whether or not it makes good physics sense to ``analytically continue'' physical spacetime into the complex realm. Wick rotation, for instance, works very well in Minkowski spacetime, and even in static spacetimes, but is more ambiguous in general spacetimes~\cite{Wick}. Similarly, the analytic continuation and complex coordinate transformations in the Newman--Janis trick (for converting Schwarzschild spacetime to Kerr spacetime, or Reissner--Nordstr\"om spacetime to Kerr--Newman spacetime) has an uncertain ontological status~\cite{Newman-Janis,Newman:1973,Rajan:2016}.
More recently, the question of physical acceptability of complex metrics has been raised in references~\cite{witten,i-epsilon}.

\item The quaternions and their variants continue to have niche applications rather than general purpose applications. Key variants include:
\begin{itemize}
	
\item Quaternions (real); extremely useful for dealing with three-dimensional rotations; but they also have many other applications~\cite{Ginibre:1965, Finkelstein:1961, Bengtsson:1987, Adler:1985a, Adler:1985b, Page:1985, Edmonds:1974};
\item Quaternions (complex); useful for dealing with Lorentz transformations~\cite{Rastall:1964,Berry:2020,Berry:2021};
\item Octonions~\cite{Borsten:2008, Chung:1987};
\item Sedonions.
\end{itemize}

\item Other algebraic systems sometimes of interest include:
\begin{itemize}
\item Clifford algebra~\cite{Crawford:1991, Greider:1984};
\item Hestenes algebra (spacetime algebra)~\cite{Hestenes:1973, 
Hestenes:1971, Hestenes:1982, Wilson:2021};
\item Grassman algebra~\cite{Zumino:1979, Galvao:1980, Arnowitt:1975}.
\end{itemize}

\item The various competing forms of non-standard analysis all have their own partisan support~\cite{Robinson,Loeb}. Some of the central constructions are:
\begin{itemize}
\item Hyper-real numbers~\cite{hyper-reals};
\item Super-real numbers~\cite{super-reals};
\item Surreal numbers~\cite{surreals1,surreals2,surreals3};
\item The Levi--Civita field~\cite{Levi-Civita}.
\end{itemize}

Of course this is moving even further afield from empirical reality (or more precisely, further from things one can think of measuring directly), so most physicists would demand very good reasons---some extremely high payoff---for moving in such a direction.

\item I should also (extremely briefly) mention the $p$-adics (and adelics). Let us just say that their direct connection to empirical reality seems to be a little strained~\cite{p-adics, Brekke:1993, Dragovich:2009, Vladimirov:1988, Dragovich:2017}. 
\end{itemize}

\item One could also side-step the rationals by using something more exotic such as a \emph{logarithmic number system} wherein the (non-zero) reals are codified as an ordered pair
$ x \longleftrightarrow \{\hbox{sign}(x), \ln |x|\}$, and zero is codified by $ 0 \longleftrightarrow \{0, \hbox{undefined}\}$. In this number system, multiplication and division are easy, but addition and subtraction are hard. 
If one approximates $ \ln |x|$ by some finite precision quantity, then this \emph{logarithmic number system} will cover a finite subset of the reals, that is not a finite subset of the rationals. 
\end{itemize}

I emphasise that one of the key distinctions we can make is between number systems of finite (but large) cardinality, countable (denumerable) cardinality, and uncountable (non-denumerable) cardinality.


\section{Discussion}\label{S:discussion}
The much discussed ``unreasonable effectiveness of mathematics'' in describing empirical reality~\cite{Wigner} is subject to  a number of significant caveats---exactly \emph{which} particular aspect of mathematics is it that is so unreasonably effective?  There are several distinct mathematical frameworks (number systems) that might plausibly be used to model empirical reality, and given the current state of affairs, physicists can still reasonably agree to differ on which is ``best''. My personal judgment of the most likely place to look is this:  
The jump from rationals to reals, though \emph{mathematically} very well motivated, might not \emph{physically} be the most appropriate step to take.  There are many number systems intermediate in strength between the rationals and the reals, and one of these intermediate number systems (most probably one of countable cardinality), might plausibly provide a better way of modelling empirical reality. 
These issues are intimately connected with the question as to whether nature is fundamentally represented by a \emph{continuum} or a \emph{discretium}. 
Physically these questions become interesting if and only if one can develop a compelling mathematical alternative to real analysis (in particular, to classical differentiation and integration) that is powerful enough to allow us to do interesting things, but (hopefully) is simple enough, and different enough, to lead to compelling new physics. For instance: Can one map the \emph{error bars} of the experimental physics into some pragmatically useful version of \emph{interval arithmetic? And how exactly should the edges of the relevant intervals be represented? Is \emph{quantum physics} really pointing toward the existence of a \emph{discretium}, or is that just an illusion? } There is a rich vein of interesting ideas here, one to which both mathematicians and  physicists could usefully contribute.

\bigskip

\hrule\hrule\hrule

\bigskip

Data availability: Not applicable

\bigskip

Conflicts of interest: The author declares no conflict of interest.

\section*{Acknowledgements}

I acknowledge support via the Marsden Fund, and via a James Cook Fellowship, both administered by the Royal Society of New Zealand.

\clearpage


\begin{thebibliography}{69}
\bibitem{marked-trisect}
Archimedes is generally credited with developing a neusis method for trisecting arbitrary angles with marked edge and compass. Various modern sources.
\bibitem{marked-cube}
Archimedes is generally credited with developing a neusis method for doubling the cube with marked edge and compass. Various modern sources.
\end{thebibliography}
\end{document}